\documentclass[12pt]{article}
\usepackage[english,german,french,polish]{babel}
\usepackage[T1]{fontenc}
\usepackage{amsfonts}

\begin{document}

\selectlanguage{english}

\baselineskip 0.73cm
\topmargin -0.4in
\oddsidemargin -0.1in

\let\ni=\noindent

\renewcommand{\thefootnote}{\fnsymbol{footnote}}

\newcommand{\SM}{Standard Model }

\newcommand{\SMo}{Standard-Model }

\pagestyle {plain}

\setcounter{page}{1}


                            
\vfill\eject

~~~~~~
\pagestyle{empty}

\begin{flushright}
IFT-- 1/16
\end{flushright}

\vspace{0.4cm}

{\large\centerline{\bf Coupling of Hidden Sector}}

\vspace{0.5cm}

{\centerline {\sc Wojciech Kr\'{o}likowski}}

\vspace{0.3cm}

{\centerline {\it Institute of Theoretical Physics, University of Warsaw}}

{\centerline {\it Pasteura 5, 02--093 Warszawa, Poland}}

\vspace{0.6cm}

{\centerline{\bf Abstract}}

\vspace{0.2cm}

A hypothetic Hidden Sector of the Universe, consisting of sterile fermions ("sterinos") and sterile mediating bosons ("sterons") of mass dimension 1 (not 2!) --- the last described by an antisymmetric tensor field --- requires to exist also a scalar isovector and scalar isoscalar in order to be able to construct electroweak invariant coupling (before spontaneously breaking its symmetry).

\vspace{0.6cm}

\ni PACS numbers: 14.80.-j , 04.50.+h , 95.35.+d 

\vspace{0.6cm}
 
\ni March 2016

\vfill\eject 

\pagestyle {plain}

\setcounter{page}{1}

\vspace{1.0cm}

\ni {\bf 1. Introduction}

\vspace{0.4cm}

In the previous decade, we have introduced a specific Hidden Sector of the Universe, consisting of sterile fermions (with mass dimension 3/2) and sterile mediating bosons (of mass dimension 1) described by an antisymmetric-tensor field $C_{\mu\nu}$ (denoted, before, by $A_{\mu\nu}$) [1,2]. In addition to the familiar structure of the Standard Model we have postulated the existence of an extra scalar isovector $ (\varphi_1 , \varphi_2 , \varphi_3)\;\;(i = 1,2,3)$ or

\begin{equation}
\varphi^+ =  \frac{\varphi_1 + i\varphi_2}{\sqrt2}\; .\; \varphi^- =  \frac{\varphi_1 - i\varphi_2}{\sqrt2}\; . \;\varphi^0 = \varphi_3 
\end{equation}

\vspace{0.1cm}

\ni and also a scalar isoscalar $\varphi_0$. While the former triplet is conserving, the latter singlet is presumed to break  spontaneously the electroweak symmetry, $<\varphi_0>_{\rm vac} \neq 0$, acting beside the popular Higgs scalar, $<h_0>_{\rm vac} \neq 0$. The introduced sterile fermions $\psi$ and sterile mediating bosons $C_{\mu\nu}$ we will call, for convenience, "sterinos" and "sterons", respectively. Then, hidden vectors $\chi_\nu$ might be named "hiddons".

If the tensor field $C_{\mu\nu}$ is gauged, $C_{\mu\nu} = {\partial_\mu} \chi_\nu - {\partial_\nu} \chi_\mu$, by a vector field $\chi_\mu$ of mass dimension 0, while field $\chi_\mu$ turns out to be absent outside $C_{\mu\nu}$, then the Lagrangian density is gauge invariant, trivially. The mass-dimension-0 vector ${\chi_\mu}(x)$ might play still a fundamental role in the creation of gravitational curvature $g_{\mu\nu}(x)$ as a specific condensate of matter: 

\begin{equation}
g_{\mu\nu}(x) = <{\chi_\mu}(x) {\chi_\nu} (x)>_{\rm con} 
\end{equation}

\ni with the Lorentz-like gauge $\,{\partial_\nu} \chi^\nu = 0\,$ and ${\partial^\nu} C_{\mu\nu} = \Box \chi_\mu$, where $\,C_{\mu\nu} = {\partial_\mu} \chi_\nu - {\partial_\nu} \chi_\mu\,$. Here, $C_{\mu\nu}$ may be represented later by the matrix (8), and a special structure of $C_{\mu\nu}$ can be investigated. If {\it e.g.}, $C^{(E)}_i = C^{(B)}_i$ $(i=1,2,3)$, then $(1/4)M^2 C_{\mu\nu} C^{\mu\nu} = 0 $ (see later on Eq. (9)) and so, tensor $C_{\mu\nu}$ is effectively massless.

\vspace{0.4cm}

\ni {\bf 2. Coupling of Hidden Sector}

\vspace{0.4cm}

The new scalars $\varphi^+\,,\,\varphi^-\,,\,\varphi^0$ and $\varphi_0$ with $<\varphi_0>_{\rm vac} \neq 0$ enable us to define electroweak-symmetry invariant coupling of Hidden Sector to the \SMo world in the following form

\begin{equation}
-\frac{1}{2}\sqrt{f}\left(\bar{\psi}\sigma^{\mu\nu} \psi + \xi\,\varphi_i\, W^{\mu\nu}_i + \eta\,\varphi_0\, B^{\mu\nu} \right) C_{\mu\nu}\,. 
\end{equation}

\ni Here, $f$ and $\xi$ or $\eta$ are masslesss unknown coupling constants. The form (3) is a subject of spontaneously electroweak symmetry breaking by $<\varphi_0>_{\rm vac} \neq 0$, in addition to the Higgs mechanism $<h^0>_{\rm vac} \neq 0$ giving Weinberg-Salam mixing

\begin{eqnarray} 
Z_\mu & = & \cos\theta_W W^0_\mu + \sin \theta_W B_\mu \, ;\nonumber \\
A_\mu & = & -\sin\theta_W W^0_\mu + \cos \theta_W B_\mu \;,
\end{eqnarray}

\ni where $Z_{\mu\nu} = \partial_\mu Z_\nu - \partial_\nu Z_\mu$ and $F_{\mu\nu} = \partial_\mu A_\nu - \partial_\nu A_\mu$. Then, if we put tentatively $\xi = \eta$, the form (3) implies (after breaking) the following neutral part of the electroweak-hidden coupling: 

\begin{equation}
-\frac{1}{2}\sqrt{f}\left[ \bar{\psi}\sigma^{\mu\nu} \psi + \xi \left( \varphi^{(F)} F^{\mu\nu}+\varphi^{(Z)} Z^{\mu\nu} \right)\right] C_{\mu\nu} \,.
\end{equation}

\ni with (valid for $\xi = \eta$ ):

\vspace{-0.2cm}

\begin{eqnarray} 
\varphi^{(Z)} & = & \cos\theta_W \varphi^0 + \sin \theta_W \varphi_0 \, ,\nonumber \\
\varphi^{(F)} & = & -\sin\theta_W \varphi^0 + \cos \theta_W \varphi_0\, . 
\end{eqnarray}

Recall that in contrast to $F_{\mu\nu} = \partial_\mu A_\nu - \partial_\nu A_\mu$, the antisymmetric tensor field does not get gauging $C_{\mu\nu} = \partial_\mu \chi_\nu - \partial_\nu \chi_\mu $ with mass dimension 2, when $\chi_\mu$ has mass dimension 0. 

Naturally, the kinetic Lagrangian density is now

\begin{equation}
-\frac{1}{4}\left[ \left( \partial_\lambda C_{\mu\nu}\right) \left( \partial^\lambda C^{\mu\nu} \right)- M^2 C_{\mu\nu}C^{\mu\nu} \right] \,. 
\end{equation}

A convenient way to represent the antisymmetric tensor field $C_{\mu\nu}$ reads as the electro\-magnetic-type matrix [2]:

\begin{equation} 
\left( C_{\mu\nu}\right)  = \left( \begin{array}{cccc}
 0 & \;\;C^{(E)}_1 & C^{(E)}_2 & C^{(E)}_3 \\
-C^{(E)}_1 & 0 & -C^{(B)}_3 & C^{(B)}_2 \\
-C^{(E)}_2 & C^{(B)}_3 & 0 & -C^{(B)}_1 \\ 
-C^{(E)}_3 & -C^{(B)}_2 & C^{(B)}_1 & 0 \end{array} \,.
\right)  
\end{equation}

\ni The trace of its square is equal to  

\begin{equation}
-\frac{1}{4}\left(C_{\mu\nu}C^{\mu\nu} \right) = \frac{1}{4}\left(C_{\mu\nu}C^{\nu\mu} \right) = \frac{1}{2} \sum_i \left({C^{(E)2}_i} -  {C^{(B)2}_i}\right) \,. 
\end{equation}

A practical formula is the square of the matrix $C_{\mu\nu}$ that becomes equal to

$$
-\frac{1}{4}\left(C_{\mu\lambda}C^{\nu\lambda}\right) = \frac{1}{4}\left(C_{\mu\lambda}C^{\lambda\nu} \right) =  
$$

\vspace{0.2cm}

\begin{small}

$$
\left( \begin{array}{cccc} \vec{C}^{(E)\,2}_1 & \left( {\vec{C}}^{(E)}\!\!\times\!\! {\vec{C}}^{(B)}\right)_1 & 
\left( {\vec{C}}^{(E)}\!\!\!\times\!{\vec{C}}^{(B)}\right)_2 & \left( {\vec{C}}^{(E)}\!\!\times\!\!{\vec{C}}^{(B)} \right)_3 \\
-\left( {\vec{C}}^{(E)}\!\!\!\times\!!{\vec{C}}^{(B)} \right)_1 & C^{(E)2}_1 \!\!-\! C^{(B)2}_2 \!\!-\! C^{(B)2}_3 & 
C^{(E)}_1 C^{(E)}_2 \!+\! C^{(B)}_1 C^{(B)}_2 & C^{(E)}_3 C^{(E)}_1\!+\!C^{(B)}_3 C^{(B)}_1 \\
-\left( {\vec{C}}^{(E)}\!\!\!\times\! {\vec{C}}^{(B)}\right)_2  &  C^{(E)}_1 C^{(E)}_2 + C^{(B)}_1 C^{(B)}_2 & 
C^{(E)2}_2 \!\!-\! C^{(B)2}_3 \!\!-\! C^{(B)2}_1 & C^{(E)}_2 C^{(E)}_3 + C^{(B)}_2 C^{(B)}_3 \\
-\left ({\vec{C}}^{(E)} \!\!\!\times \!{\vec{C}}^{(B)} \right)_3 & C^{(E)}_3 C^{(E)}_1 + C^{(B)}_3 C^{(B)}_1 & 
C^{(E)}_2 C^{(E)}_3 \!+\! C^{(B)}_2 C^{(B)}_3 & C^{(E)2}_3 \!\!-\! C^{(B)2}_1\!\!\!-\! C^{(B)2}_2  \end{array} \right),  \eqno{(10)}     
$$

\end{small}

\addtocounter{equation}{+1}

\vspace{0.4cm}

\ni where ${\vec{C}}^{(E)} = \left(-{C^{(E)}_i} \right)$ and ${\vec{C}}^{(B)} = \left(-{C^{(B)}_i} \right)$ $\;(i = 1,2,3)$.

Note that the mass term $(1/4) M^2 \left( C_{\mu\nu}C^{\mu\nu} \right) = 0$ of the field $\,C^{\mu\nu}\,$ vanishes, when $\sum_i {C^{(E)2}_i} = \sum_i {C^{(B)2}_i}$ . Here, $C^{(E)}_i$ and $C^{(B)}_i$ are intrinsic degrees of freedom for a "steron" described by the field $C_{\mu\nu}$ (in analogy to the spin of a "sterino" or of another Dirac bispinor field).

\vspace{0.4cm}

\ni {\bf 3. Fermionic {\it versus} bosonic coupling}

\vspace{0.4cm}

In particle physics a fundamental role is played by 16 independent Dirac matrices building up 5 Lorentz covariant forms:

\vspace{0.4cm}

\begin{eqnarray} 
S_{\mu\nu} & \equiv & \;\; \frac{1}{2}\left\{\gamma_\mu, \gamma_\nu\right\}\;\; \equiv\;\; g_{\mu\nu}\,,\; S^{(p)}\;\equiv\, \gamma_5 \;\:\, \equiv \;\; \gamma_0 \gamma_1 \gamma_2 \gamma_3  \,,\, \nonumber \\
V_\mu\: & \equiv & \gamma_\mu = \left\{\begin{array}{crr} 0 & {\rm for} & \mu = 0 \; \\ \beta \alpha_k & {\rm for} & \mu = k \end{array} \right. \,,\hspace{0.39cm}V^{(p)}_\mu \equiv \:\:\gamma_\mu \gamma_5 \,, \nonumber \\
T_{\mu \nu}\! & \equiv & \sigma_{\mu\nu} = \frac{i}{2}[\gamma_\mu,\gamma_\nu] = \left\{\begin{array}{crr}  i \alpha_l & {\rm for} & \mu = 0\,,\,\nu = l \;  \\ \varepsilon_{klm} \sigma_m & {\rm for} & \mu = k \,,\,\nu=l \\ -i \alpha_k & {\rm for} & \mu = k \,,\,\nu= 0 \\\end{array} \right. \,, 
\end{eqnarray}

\vspace{0.2cm}

\ni where $\left\{\gamma_\mu, \gamma_\nu\right\} = 2g_{\mu\nu}\;,\; (\mu,\nu = 0,1,2,3)$ and $\left[\sigma_k\,,\,\sigma_l \right] = 2i \varepsilon_{klm} \sigma_m\,,\,(k,l,m = 1,2,3)$.

These covariant forms determine couplings of mediating bosons (with mass dimension 1) to fermionic pairs (with mass dimension 3/2 + 3/2 = 3). The mass dimension of interaction Lagrangian density is then 1 + 3 = 4, while kinetic Lagrangian density of mediating bosons gets the dimension 2 + 2 = 4.

For instance, electrons and photons are coupled according to the electromagnetic interaction Lagrangian density. 

\begin{equation}
e \bar\psi\, \gamma_\mu \psi\, A^\mu . 
\end{equation}
 
\ni Here, $A_\mu$ and $F_{\mu\nu} = \partial_\mu A_\nu - \partial_\nu A_\mu$ have mass dimension 1 and 2, respectively, while $-(1/4) F_{\mu\nu}F^{\mu\nu} = (1/2) \sum_k{(E^2_k - B^2_k)}$ carries  mass dimension 4 for the kinetic Lagrangian density of photons.
 
All other fermionic couplings among those given by Eqs. (11) are also realized experimentally, except for the Pauli-type antisymmetric tensor coupling 

\begin{equation}
\frac{1}{2} \sqrt{f\,} \bar\psi\, \sigma_{\mu \nu} \psi\, C^{\mu \nu} \,, 
\end{equation}
 
\ni where a new antisymmetric tensor field is introduced (see Eq. (8)) :

\begin{equation}
\left( C_{\mu\nu}\right)  = \left\{ \begin{array}{ccc}
 C^{(E)}_l & {\rm for} & \mu = 0 \,,\, \nu = l \,, \\
 \varepsilon_{klm}{C_m^{(B)}} & {\rm for} & \mu = k \,,\, \nu = l \,, \\
-C^{(E)}_k & {\rm for} & \mu = k \,,\, \nu = 0 \, \\ \end{array} \right.
\end{equation}

\ni $(k,l,m = 1,2,3)$. The mathematical existence of this tensor (isoscalar) suggests an experimentally new scalar isovector and scalar isoscalar.

Arguing for extending fermionic tensor coupling (13) to bosonic tensor coupling, we obtain Eq. (3) as an electroweak-symmetry invariant coupling of Hidden Sector to the \SMo world (before spontaneously breaking the symmetry).

In Ref. [2] we considered, for illustration, the process

\begin{equation} 
C C \rightarrow \varphi^{(F)}_{\rm phys} \gamma\, \varphi^{(F)}\!_{\rm phys} \gamma \rightarrow \gamma \gamma 
\end{equation}

\ni (here, we use the notation $C$ instead of $A$).

\vfill\eject

\vspace{1.5cm}  

{\large\centerline{\bf Appendix A}}

\vspace{0.3cm}

{\centerline {\bf Free steron {\it versus} sterino}}

\vspace{0.4cm}  

With the interaction and kinetic parts of steron Lagrangian density, Eqs. (3) and (7), the Lagrangian field equation for $C^{\mu\nu}$ reads:

$$
(\Box - M^2) C^{\mu\nu} \equiv -\sqrt{f}\left( \bar{\psi} \sigma^{\mu\nu} \psi +\xi \varphi_i W^{\mu\nu}_i +  \eta \varphi_0 B^{\mu\nu} \right) \,,  \eqno{(\rm A\,1)}
$$

In the limit of $f \rightarrow 0$, we obtain a free equation for $C^{\mu\nu}$, $(\Box - M^2) C^{\mu\nu} = 0$, getting the plane-wave solutions:

$$
C^{\mu\nu}_a(x) = \frac{1}{(2\pi)^{3/2}} \frac{1}{\sqrt{2 k_0}} e^{\mu\nu}_a e^{-ik\cdot x}  \,,  \eqno{(\rm A\,2)}
$$

\ni where

$$
k_e = \sqrt{\vec{k}^2 + M^2 }  \eqno{(\rm A\,3)}
$$

\ni and

$$
\left(e^{\mu\nu}_a \right) = \left(e^{\mu\nu}\right)_a = \left(\begin{array}{cccc} 0 & e^{(E)}_1 & e^{(E)}_2 & e^{(E)}_3 \\ -e^{(E)}_1 & 0 & -e^{(B)}_3 & e^{(B)}_2 \\ -e^{(E)}_2 & e^{(B)}_3 & 0 & -e^{(B)}_1 \\ -e^{(E)}_3 & -e^{(B)}_2 & e^{(B)}_1 & 0 \end{array} \right)_{\!\!a} \eqno{(\rm A\,4)}
$$

\vspace{0.3cm}

\ni due to Eq. (8). Here, three independent polarizations $ \vec{e}^{\,(E,B)}_a = \left( e^{(E,B)}_k\right)_a $ ($a = 1,2,3$ and $k = 1,2,3$) are chosen ortonormal, separately for $E$ and $B$ :

$$
\vec{e}^{\,(E,B)}_a \cdot \vec{e}^{\,(E,B)}_b = \delta_{ab} \;\;\;,\;\;\; {\sum^3_{a=1} {e}^{(E,B)}_{k a} {e}^{(E,B)}_{l a} = \delta_{k l}} \,. \eqno{(\rm A\,5)}
$$

Now, we may impose {\it a priori} a hypothetic relation between polar and axial polarizations, $\vec{e}^{\,(E)}_a$ and $\vec{e}^{(B)}_a$, within the solution (A 2), putting the constraint

$$
\vec{e}^{\,(E)}_{1,2,3} \times \vec{e}^{\,(E)}_{3,1,2} = \vec{e}^{(B)}_{2,3,1} \;, \eqno{(\rm A\,6)}
$$

\ni besides the identity:

$$
\vec{e}^{\,(E)}_{1,2,3} \times \vec{e}^{\,(E)}_{2,3,1,} = (+ \;\,{\rm or}\;\,-) \vec{e}^{\,(E)}_{3,1,2} \eqno{(\rm A\,7)}
$$

\ni (in the right- or left-handed frame of reference, respectively). Then, the constraint (A 6) can be presented trivially as

$$
\vec{e}^{\,(B)}_a = (+ \;\,{\rm or}\;\,-) \vec{e}^{\,(E)}_a \eqno{(\rm A\,8)}
$$

\ni and so,

$$
\vec{e}^{\,(B)2}_a = \vec{e}^{\,(E)2}_a \eqno{(\rm A\,9)} 
$$

\ni $(a = 1,2,3)$. Thus, from Eqs. (A 4) and (A 9) the trace of matrix $e^{\mu\nu}_a$ squared (minus) is

$$
e_{\mu\nu\,a} e^{\mu\nu}_a = 2 \left( \vec{e}^{\,(B)2}_a - \vec{e}^{\,(E)2}_a \right) = 0\,, \eqno{(\rm A\,10)}
$$

\ni the last step being valid, when one uses the constraint (A 9). Then, we can  infer that the effective mass term $(1/4) M^2 C_{\mu\nu} C^{\mu\nu}$ of the steron field $C^{\mu\nu}$ vanishes (see Eq. (7)).

In a similar way, we get the sterino Lagrange field equation:

$$
\left(\gamma^\mu i \partial_\mu - \frac{1}{2}\sqrt{f\,}\, \sigma^{\mu\nu} C_{\mu\nu} - m_\psi \right)\psi = 0 ,\eqno{(\rm A\,11)}
$$

\ni when we apply the energy and kinetic parts of sterino Lagrangian density, Eq. (3) and the term

$$
\
\bar{\psi}\left(\gamma^\mu i \partial_\mu - m_\psi \right) \psi   \,, \eqno{(\rm A\,12)}
$$

\ni respectively. Due to the identity

$$
\frac{1}{2} \sigma^{\mu\nu} C_{\mu\nu} = i\, \vec{\alpha}\cdot \vec{C}^{\,(E)} + \vec{\sigma}\cdot \vec{C}^{\,(B)}   \eqno{(\rm A\,13)}
$$

\ni following from the formulae (8) and (11), we can rewrite the sterino field equation (A 11) as

$$
\left[\gamma^\mu i \partial_\mu - m_\psi - \sqrt{f}\left( i\vec{\alpha}\cdot \vec{C}^{\,(E)} + \vec{\sigma}\cdot \vec{C}^{\,(B)} 
\right)\right] = 0 \,, \eqno{(\rm A\,14)}
$$

\ni where $ \vec{C}^{\,(E,B)} = \left( C^{\,(E,B)}_k \right) $ .

A physically interesting case might be  a sterino $\psi(x)$ embedded in the uniform steron field $ \vec{C}^{\,(E)} = \overrightarrow{\rm const} = (0,0,C) $ and $ \vec{C}^{\,(B)} = \overrightarrow{\rm const} = (0,0,C') $. Then, from Eq. (A 14) we infer that

$$
\left[ E - \vec{\alpha}\cdot\vec {p} - \beta m_\psi - \sqrt{f}\, \gamma_3\left(i C + \gamma_5 C'\right)\right] \psi = 0 \,, \eqno{(\rm A\,15)}
$$

\ni for $i \partial_\mu \psi(\vec{p}) = p_\mu \psi(\vec{p})$ with $\vec{p}$ denoting the momentum of sterino, while $E(\vec{p})$ is its energy spectrum, Multiplying Eq. (A 15) on the lhs by $\left[E + \alpha\cdot\vec{p} + \beta m_\psi + \sqrt{f} \gamma_3 (iC + \gamma_5 C')\right]$, we get sterino quadratic spectrum:

$$
\left[ E^2 - \vec{p}^{\;2} - m_\psi^2 - f(C^2 + C'^2)- 2\sqrt{f\,} C'( \sigma_3 m_\psi+ i \gamma_1 p_2 - i \gamma_2 p_1 )\right]\psi = 0 \,.  \eqno{(\rm A\,16)}
$$

We solve this spectrum in terms of sterino momenta $p_1, p_2, p_3$, finding the eigenvalues of the complete set of independent observables:

\vspace{-0.2cm}
 
$$
\frac{1}{2} \sigma_3 \psi = m_s \psi \equiv \pm\frac{1}{2} \psi  \eqno{(\rm A\,17)} 
$$

\ni and

\vspace{-0.2cm}
 
$$
(i\gamma_1 p_2 - i\gamma_2 p_1)\psi = \pm\sqrt{(i\gamma_1 p_2 - i\gamma_2 p_1)^2\,}\, \psi \equiv \pm \sqrt{{p^2}_1 + {p^2}_2\,}\, \psi  \eqno{(\rm A\,18)}
$$

\ni with

\vspace{-0.2cm}
 
$$
E^2 = {\vec{\,p}}^2 + {m_\psi}^2 +f(C^2 + {C'}^2)+2\sqrt{f\,}C'\left(\pm m_\psi \pm\sqrt{{p^2}_1 + {p^2}_2\,}\right)\;.  \eqno{(\rm A\,19)}
$$

\ni Here, $\vec{\gamma} = \beta\, \gamma_5\,\vec{\sigma} = \beta \vec{\alpha}\,,\, (i \vec{\gamma})^\dag = i \vec{\,\gamma}\,$ and $\,(i\gamma_1)^2 = {\bf 1} =  (i\gamma_2)^2$ as well as $\{i\gamma_1,i\gamma_2\} = 0$. Thus,

\vspace{-0.2cm}
 
$$
\left[\sigma_3\,,\{i\gamma_k\,,\,i\gamma_l\}\right] = 2\delta_{kl}[\sigma_3\,, {\bf 1}]= 0  \eqno{(\rm A\,20)}
$$

\ni so, squares of $i\gamma_1$ and $i\gamma_2$ are independent of $\sigma_3$. 

\vspace{1.5cm}  

\vfill\eject

{\large\centerline{\bf Appendix B}}

\vspace{0.3cm}

{\centerline {\bf Maxwell's Hidden Equations}}

\vspace{0.4cm}

The streron field $C_\mu\nu$ can be split into electro- and magnetic-type parts (see [3]) according to Eq. (8). Here, the trace of its square is given in Eq. (9). In the case of {\it e.g.}, ${C^{(E)2}_k} = {C^{(B)2}_k}\;(k = 1,2,3)$, the effective steron masss $(1/4)M^2 C_{\mu\nu}C^{\mu\nu}$ vanishes.

When sterons are gauged.

$$
 C_{\mu\nu} = \partial_\mu \chi_\nu - \partial_\nu \chi_\mu\;\;\;{\rm or}\;\;\;\;\, \begin{array}{l} \vec{C}^{(E)} = -\partial_0\, \vec{\chi} - \vec{\partial} \,\chi_0 \,, \\ \vec{C^{((B)}} = {\bf rot} \vec{\chi} \end{array}\,. \eqno{(\rm B\,1)}
$$

\ni with $C_{\mu\nu}$ and $\chi_\mu$ of the mass dimension one and zero, respectively, then:


\begin{eqnarray*} 
{\bf rot}\,\vec{C}^{(E)} + \partial_0\,\vec{C}^{(B)} & = & 0 \,, \\
{\rm div}\, \vec{C}^{(B)} & = & 0\,. 
\end{eqnarray*}

\vspace{-1.7cm}

\begin{flushright}
$(B\,2)$
\end{flushright}

On the other hand, the Lorentz-type gauge condition 

$$
\partial^\nu\chi_\nu = 0  \eqno{(\rm B\,3)}
$$

\ni implies from (7)

$$
\partial^\nu C_{\mu\nu} = \Box\chi_\mu \, \eqno{(\rm B\,4)}. 
$$

Now, from Eqs, (10) and (5), we obtain

$$
\Box \chi_\mu  = \left\{ \begin{array}{ccc}
{\rm div}\, \vec{C}^{(E)} & {\rm for} & \mu = 0 \,, \\
\left({\bf rot}\,\vec{C}^{(B)} - \partial_0\, \vec{C}^{(E)} \right)_\mu & \;\;{\rm for}\;\; & \mu = 1,2,3\,,\,  \\ \end{array} \right. \eqno{(\rm B\,5)}
$$

\ni or


\begin{eqnarray*}  
{\bf rot}\,\vec{C}^{(B)} - \partial_0\,\vec{C}^{(E)} & = & \Box\vec    {\chi} \,, \\
{\rm div}\, \vec{C}^{(B)} & = & \Box \chi_0\,. 
\end{eqnarray*}

\vspace{-1,72cm}

\begin{flushright}
$(B\,6)$
\end{flushright}

\ni We can see that $\Box \chi_0$ and $\Box \vec{\chi}$ are time-like and space-like sources of sterons $C_{\mu\nu}$. Both of them form together the "continuity equation":

$$
\Box \left(\partial_0\chi_0 - \vec{\partial}\vec{\chi}\right) = \Box\left(\partial^\nu \chi_\nu \right)= 0 \;\;\;\;{\rm or}\;\;\;\; \partial_0\, \rho + {\rm div}(\rho\, \vec{v}) = 0 \,, \eqno{(\rm B\,7)}
$$

\ni where

$$
\rho = \Box \chi_0 \;\;,\;\; \rho\, \vec{v} = -\Box \vec{\chi}\,. \eqno{(\rm B\,8)}
$$

\ni Here, $\vec{v}$ is the field velocity of hiddons being antinormally orientated sources of sterons.

The four formulae (8) and (12) are Maxwell-type equations, acting on sterons $C_{\mu\nu}$ with mass dimensions one ("Maxwell's hidden equations"), defining "hidden electromagnetism". In the world of {\SM}, electroweak symmetry is actually active ({\it plus} hypothetic new scalar isovector and scalar isoscalar fields recently intoduced [1,2,3]).

There is also a cross-coupling betwen the Hidden Sector and \SMo world, {\it cf.} Eq. (2), the latter being electroweak-symmetry invariant (before spontaneously breaking the electroweak symmetry). As a result of the coupling (2), the conventional Maxwell's equations (already electroweakly unified) become extended ("Maxwell's extended equations"). ({\it Cf.} Eq. (10) in Ref. [2]).


\vspace{1.5cm}

{\large{\centerline{\bf References}}}

\vspace{0.4cm}

\baselineskip 0.73cm

{\everypar={\hangindent=0.65truecm}
\parindent=0pt\frenchspacing

{\everypar={\hangindent=0.65truecm}
\parindent=0pt\frenchspacing

[1]~W.~Kr\'{o}likowski, {\it Acta Phys. Polon.} {\bf B 40}, 2767 (2009). 

\vspace{0.2cm}

[2]~W.~Kr\'{o}likowski, {\it Acta Phys. Polon.} {\bf B 41} 1277 (2010),

\vspace{0.2cm}

[3]~Cf. {\it e.g.} L.D.Landau and A.E.M.Lipshitz, {\it The Classical Theory of  Field}, Pergamon Press, New York (1977). 

\vspace{0.2cm}

\vfill\eject

\end{document}